\documentclass{article}
\usepackage{calor2kcepro}
\usepackage{graphicx}
\begin{document}
\title{ 
  PERFORMANCES OF THE NA48 LIQUID KRYPTON CALORIMETER
  }
\author{
  Guillaume Unal\thanks{\ on behalf of the NA48 collaboration 
  (Cagliari, Cambridge, CERN, Dubna, Edinburgh, Ferrara, Firenze,
   Mainz, Orsay, Perugia, Pisa, Saclay, Siegen, Torino, Warsaw, Wien)}        \\
  {\em LAL, IN2P3-CNRS et Universit\'e Paris-Sud, BP.34, 91898 Orsay, France}\\
  }
\maketitle
\baselineskip=11.6pt
\begin{abstract}
  The NA48 experiments aims at a precise measurement of direct
  CP violation in the neutral Kaon system. This puts
  stringent requirements on the electromagnetic calorimeter used
  to detect photons of average energy 25 GeV.
  The choice of NA48 is a quasi homogeneous Liquid Krypton
  calorimeter with fast readout. The operation of this
  device and the performances achieved are described.
\end{abstract}
\baselineskip=14pt
\section{Introduction}

 The NA48 experiment at the CERN SPS aims at a precise
measurement of the $\epsilon'/\epsilon$ parameter,
characterising direct CP violation in the
neutral kaon decays. This parameter can be accessed
experimentally through the double ratio $R$ of the decay
rates :
\begin{equation}
R = \frac{ \Gamma( K_L \rightarrow \pi^0 \pi^0) /  \Gamma( K_S \rightarrow 
\pi^0 \pi^0)}{ \Gamma( K_L \rightarrow \pi^+ \pi^-) / \Gamma( K_S \rightarrow \pi
^+\pi^-)}  \approx  1 - 6 \times Re(\epsilon'/\epsilon) \nonumber \\
\end{equation}
 To achieve the required accuracy of $\approx$ $2\times10^{-4}$ on
$\epsilon'/\epsilon$, a large statistic is needed.
To minimise systematic
effects, NA48 collects the four modes
at the same time using simultaneous
$K_L$ and $K_S$ beams and from the same decay region, such that
most effects cancel at first order in the double ratio.
A description
of the analysis of the data taken in 1997 
can be found in~\cite{na48_epsilon}.

 The $\pi^0 \pi^0$ decay mode is identified by measuring
four photons in the calorimeter, in the energy range 3-100 GeV,
with an average energy of $\approx$~25~GeV. The calorimeter
is located $\approx$~100 m downstream of the decay region.
Photon angles are not directly measured and therefore
the distance $D$ between the kaon decay point and the calorimeter
is computed assuming the Kaon mass for the four photons, and
using the measured photon energies and positions transverse
to the beam axis :
\vspace{1mm}
\begin{eqnarray}
D & = & \sqrt{\Sigma_{i>j} E_i E_j [ (x_i-x_j)^2 + (y_i-y_j)^2]} / M_K \nonumber
\end{eqnarray}
\vspace{1mm}
The fiducial decay region is defined by applying cuts on the
reconstructed value of $D$. This decay region should be defined
identically for neutral and charged decays and this implies
a good control of systematic effects in measuring
photon energies and positions.
The invariant masses of the two photon pairs
are then computed using $D$ and compared to the nominal $\pi^0$ mass to
select $K_{L,S} \rightarrow \pi^0 \pi^0$ candidates. 
To achieve a good rejection of the $K_L \rightarrow 3 \pi^0$
background (whose decay rate is $\approx$ 200 times higher
than the CP violating $K_L \rightarrow \pi^0\pi^0$ mode), a
$\pi^0$ mass resolution of $\approx$ 1 MeV is required.
In addition, NA48 is using proton tagging
to distinguish $K_L$ and $K_S$ decays : the time of each
proton directed towards the $K_S$ production target
is recorded and compared to the decay time measured in the
detector. If a coincidence is found the event is classified
as $K_S$, otherwise it is classified as $K_L$. This requires
a very reliable measurement of the event time with a good
(better than 500~ps) resolution.
 
 The requirements on the performances of the calorimeter
can be summarised as follows : \\
1) Energy resolution : should be better than 1\% at 25~GeV
and above, with a good uniformity over the
calorimeter (long range variations not exceeding $\approx$~0.1\%) \\
2) Position resolution : of the order of 1 mm \\
3) Time resolution : better than 500~ps \\
4) Non linearity : residual non linearities should
 be understood at the level of 0.1\% in the energy range 3-100~GeV \\
5) Stability : The calorimeter should give a stable
 response over several years, in a high rate of Kaon decays
 (the $K_L$ decay rate seen by the detector is $\approx$ 500~kHz)

 To fulfil these goals, NA48 has decided to use
a quasi homogeneous Liquid Krypton calorimeter,
operated as ionization calorimeter. This calorimeter
is almost fully active, which ensures a very good
intrinsic energy resolution. The use of cold noble liquid
allows a very good stability of the response. Initial current readout
technique with a fast shaping allows to operate
in a high rate environment and to achieve a good time
resolution.

 Table \ref{table1} summarises the characteristics of noble
liquids. Clearly, liquid Argon is not dense enough
for a homogeneous calorimeter.  Xenon would offer a smaller radiation length than Krypton, but
its Moliere radius, giving the transverse size
of showers, which is in our case a more important
parameter, is in fact close to the one of the Krypton.
Liquid Krypton was therefore chosen as medium.
 The calorimeter is briefly described in the next section.
The reconstruction of pulses and showers are then
described with the performances achieved, with
emphasis on the energy resolution and the linearity.

\begin{table}[h]
\centering
\caption{ \it Summary of noble liquid characteristics.}
\begin{tabular}{l|cccc}
 & Z & density (g/cm$^3$) & X0 (cm) & R(Moliere)\cite{ref_mol} (cm) \\
\hline
Ar & 18 & 1.4  & 14.0 & 9.2 \\
Kr & 36 & 2.4  & 4.7 & 6.1 \\
Xe & 54 & 3.0  & 2.9 & 5.5 \\
\hline
\end{tabular}
\label{table1}
\end{table}


\section{The Liquid Krypton calorimeter}

\subsection{Calorimeter structure}

\begin {figure}[ht]
\begin{center}
\parbox{0.45\linewidth}
{
\hspace{-0.5cm}
\includegraphics*[scale=0.50]{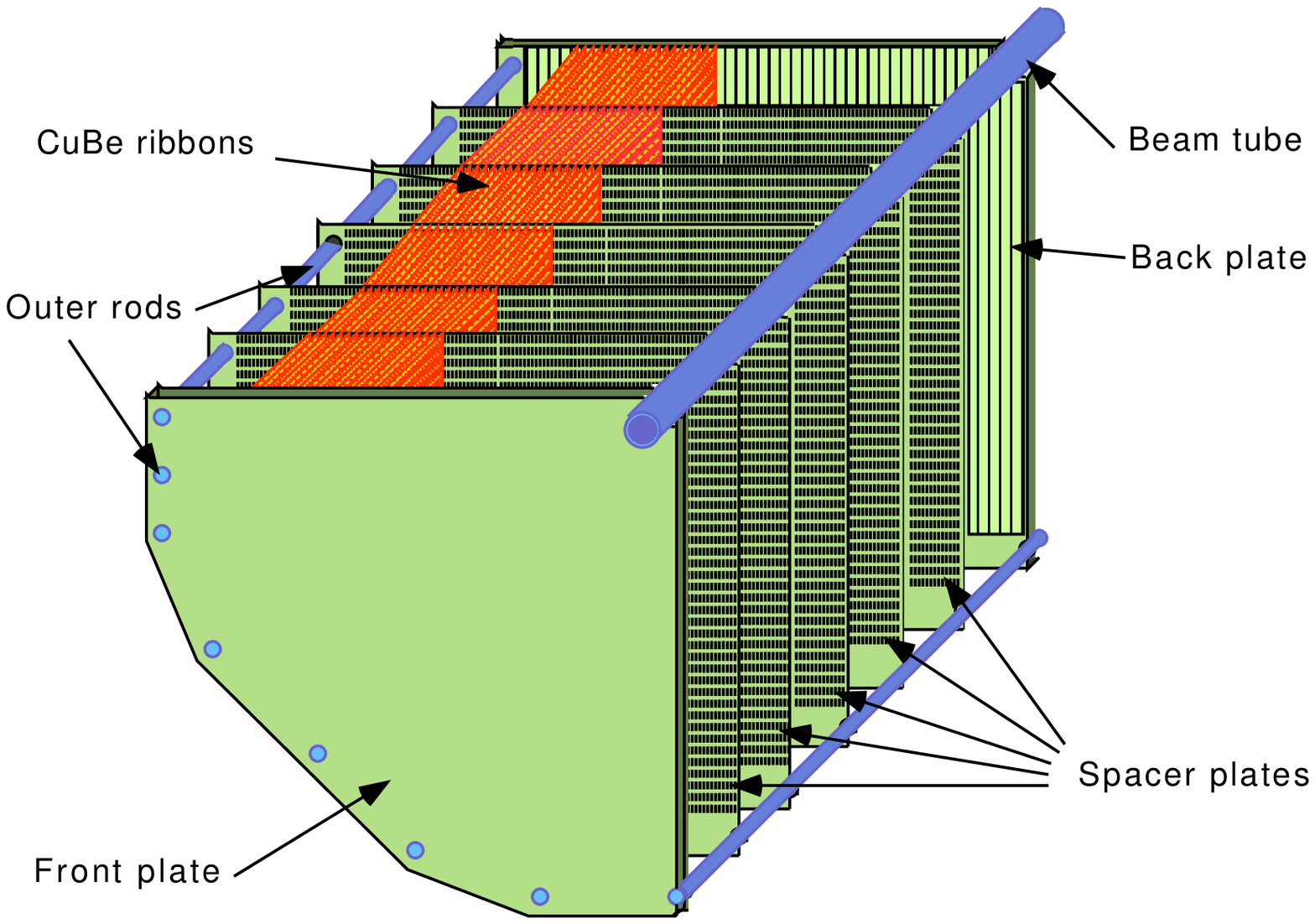}
}
\parbox{0.45\linewidth}
{
\hspace{2.5cm}
\includegraphics*[scale=0.27]{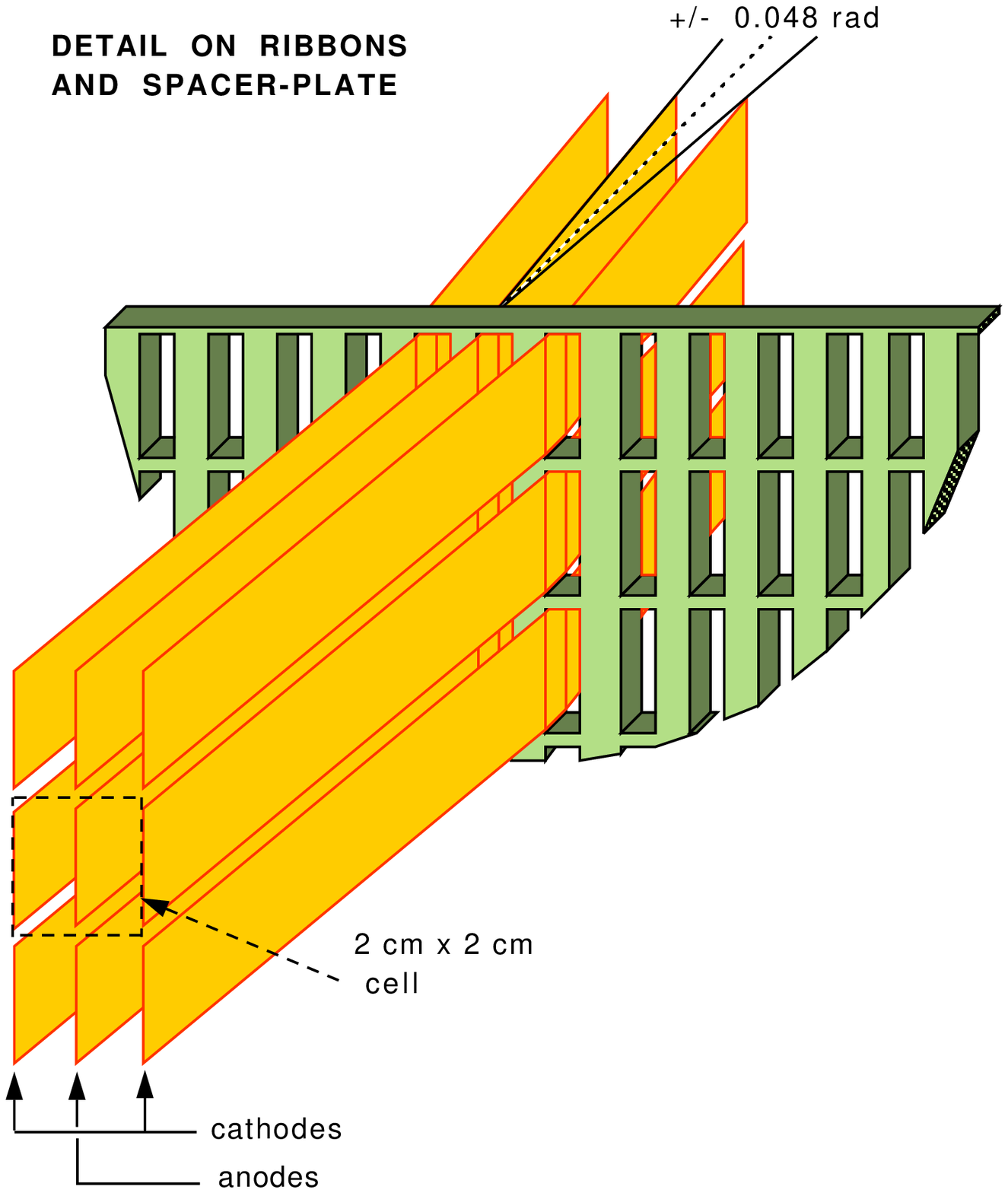}
}
\caption{\it Electrode structure of one quarter of the
calorimeter (left) and details on ribbons and one spacer
plate (right).}
\label {fig_lkr1}
\end{center}
\end {figure}

 The calorimeter is made of a bath of 
$\approx$~10~m$^3$ of liquid krypton at 120~K with
a total thickness of 125~cm ($\approx$ 27 radiation lengths)
and a octagonal shaped active cross-section of $\approx$ 5.5~m$^2$.
 A 8~cm radius vacuum tube goes through the calorimeter
to transport the undecayed neutral beam.
 Thin Copper-Beryllium ribbons (of
dimensions 40$\mu$m $\times$ 18 mm $\times$ 127 cm) stretched
between the front and the back of the calorimeter form
a tower structure readout. The 13212 readout cells
each have a cross-section of $\approx$ 2$\times2$ cm$^2$ and
consist (along the horizontal direction)
of a central anode (at the high voltage)
in the middle of two cathodes (kept at the ground). The
gap size $d$ is therefore $\approx$ 1 cm. There
is a 2~mm vertical separation between the electrodes. The drift
of the ionization electrons towards the anode produces
a current $i$ which in the limit of uniform ionization
across the gap and constant electric field can be written
as~:
\begin{equation}
i(t) = q \times \frac{v_d}{d} \times (1 - \frac{t}{t_d}) \nonumber
\end{equation}
where $q$ is the charge one wants to measure, $v_d$
the drift velocity
\footnote{typically 3100 m$\cdot$s$^{-1}$ for a high voltage of 3000 V}
and $t_d$ the total drift time
across the gap\footnote{typically 3.2$\mu$s}. 
To guarantee a good uniformity of
the initial current ($i(t=0)$), a good accuracy
of the gap is required. This is enforced by the use
of five precisely machined spacer plates (fibreglass 
reinforced epoxy, 5mm thick, $\approx$~0.025 X0),
normal to the detector axis and located every 21 cm.
The spacer plates guide the ribbons in a accordion
geometry with a $\pm$~48~mrad zig-zag angle, in
the horizontal direction. Configurations
where the shower core is aligned with an
electrode, resulting in a drop of the response,
are thus avoided.
The accuracy achieved in the position
of the holes in the spacer plates and therefore
in the gap size is typically
$\pm$ 45 $\mu$m. The overall tower structure
is projective towards the middle of the Kaon decay
region, which is located $\approx$ 114 m upstream
of the calorimeter. The measurement of photon
positions becomes thus independent of fluctuations
in the longitudinal shower development. Because
of the projectivity, the gap size is increasing
by 1\% between the front and the back of the calorimeter.
Figure \ref{fig_lkr1} shows the electrode structure
of one quarter of the calorimeter.

 The calorimeter is housed in a vacuum insulated
cryostat. The warm windows are made of aluminium foil
4mm thick followed on the beam entrance side by a cold
double walled stainless window. The outer wall which
holds the pressure is convex and 2.9~mm thick, the inner
is flat and only 0.5~mm thick. A light
honeycomb aluminium structure
connects the two foils. The total amount of matter
before the beginning of the liquid krypton volume
is $\approx$ 0.8 X0 including all the NA48 detector
elements upstream of the calorimeter, out of which
$\approx$ 0.63 X0 comes from the cryostat and the
calorimeter itself.
 Krypton evaporates continuously at the top of the cryostat
and flows through an external purifier to be subsequently
recondensed and returned to the cryostat. 
The krypton purity is such that the lifetime of secondary
electrons exceeds 100~$\mu$s ( to be compared to a shaping
time of $\approx$ 70 ns). Long term temperature variations
within the calorimeter are $<$~$\pm$0.3~K, with 
negligible influences on the calorimeter performances.

\subsection{Electronics}

 Preamplifiers are located on the back plane
of the electrode structure. A dual
hybrid version of cold charge integration preamplifier,
of BNL type using silicon JFET technology, was designed
\cite{ref_pa}.
The restoring time chosen is 150 ns. The output signal
from the preamplifiers travels in 6~m long coaxial cables
to the feed-throughs. Transceiver boards \cite{ref_tx} are mounted
directly on the warm side of the feed-throughs. They
perform pole-zero compensation (to restore the
quasi ideal triangular shape of the signal, with a
rise time of $\approx$ 22 ns), amplification and drive
differential output to the digitising electronics via 10~m
long shielded twisted pairs cables. The last stage
of the electronics \cite{ref_cpd} performs the final shaping of the
signal using a Bessel filter. The shape
of the pulse after this filter is almost symmetrical
with a width of 70~ns, risetime of~43 ns and a fall time
of 52~ns, with an undershoot of $\approx$-3\% of the
maximum lasting during the total
drift time ($\approx$ 3.16 $\mu$s). This signal is fed
to a gain switching amplifier, where the gain choice
is based on the signal before shaping, and then to a 
10 bit 40 MHz flash ADC. The event times are
asynchronous with the clock frequency.
Four gains are used with
relative values of 1,2.86,6.89 and 16.80, allowing
to cover the energy range $\approx$ -1.5 GeV to 55 GeV
per channel. In the highest gain, the sensitivity is
$\approx$ 3.5 MeV/count. The total electronic noise
per channel (dominated by the preamplifier noise) is
$\approx$ 10 MeV. The total dispersion of the gains
of the electronic chain from channel to channel
is around 3\%. The shaped pulse is also used
as input of the trigger system \cite{ref_trig}.

 The response uniformity and stability of the
full electronic chain is controlled
by a calibration system. Calibration pulses are produced
locally in the liquid Krypton so as not to suffer from
distortion or cross-talk dues to long transmission lines.
The generator produces a step of injected current
followed by an exponential decay. The amplitude
of the signal is given by a reference voltage $V$ controlled
by a 15 bit DAC. The injected current can be written
as $i$ = $\kappa \times V$, where the constant $\kappa$ depends
on the individual components of the calibration circuit.
The intrinsic dispersion of $\kappa$ is several percent.
 To achieve a good calibration accuracy, $\kappa$ was
measured for each channel on a test bench, by equalising
the calibration signal, produced for a given DAC value,
to the signal obtained by directly injecting a calibrated
current. The injected current is generated via an
external high precision resistor. The ratio of
these two calibrations gives the calibration constant $\kappa$,
expressed in mA/V.
 These constants were measured at the level
of $\approx$  1\% accuracy, for each channel, at room
temperature and at 120 K. The technique used simulates the
behaviour of the real physics pulses, including the
attenuation due to the HV decoupling capacitor, the 
capacity of the cell, and the capacitive coupling
of adjacent cells.

\subsection{Calorimeter Operation}

 The calorimeter assembly has been completed in 1996 when
test data with part of the readout electronic were taken.
The calorimeter was fully operational in 1997 for
the first data taking period for the $\epsilon'/\epsilon$
measurement. During these two years, the calorimeter had
to be operated at a high voltage of 1500 V, lower
than expected, because of problems with a small fraction
of the blocking capacitors. This led to a small
space charge effect from accumulation of positive ions
in the gap, distorting the electric field \cite{ref_space}.
In the winter 1997/1998, all the blocking capacitors were
replaced. This allowed to operate the calorimeter at a
high voltage of 3000 V, reducing the space charge effect
to a negligible level, and allowing a noise reduction
of $\approx$ 25\% (thanks to the higher drift velocity).
In 1998,1999 and 2000, data were taken with this configuration.
 Typically, there are $\approx$ 50 faulty channels
(out of $\approx$ 13000 in the calorimeter). 30 of those are
dead preamplifiers in the liquid krypton. This small mortality
happened when the calorimeter was cooled down in 1998. This
number has been very stable since, the calorimeter
being always kept cold and full of krypton.

\section{Pulse reconstruction}

 For each channel readout, the information of 10 ADC
samples is available. The timing is adjusted such
that the first two samples are before the beginning of the
triggered signal. They can therefore be used to perform
a baseline check to determine if the pulse sits on the
undershoot of a out-of-time shower. If this is the case,
the corresponding baseline is subtracted. The gain
choice is done typically 2 time samples before
the maximum of the signal and the same gain is kept
for five or more time samples. Thus usually, the
same gain value is used for all the samples needed to reconstructed
the pulse.
 The  pulses are converted to energy using a linear relation 
$s$ = $g_{gain}$ $\times$ (ADC-Offset$_{gain}$) derived from calibration
events. From this calibrated pulse, the energy ($E$) and
time ($T$) of the signal are reconstructed 
using a digital
filter method~:
\begin{eqnarray}
E & = & \Sigma a_i \times s_i \nonumber \\
T & = & \frac{1}{E} \Sigma b_i \times s_i
\end{eqnarray}
where $s_i$ is the sampled calibrated pulse, and $a_i$ and $b_i$
are digital filter coefficients, derived from the
measured pulse shape in calibration events.
These
coefficients are binned as a function of the time of the
signal, and one iteration is performed to measure
both $E$ and $T$.
 To improve the accuracy of the procedure, the channels
are divided into 10 categories according to the
observed pulse width, and coefficients are computed
for each category. In calibration events, the
pulse reconstruction accuracy is $\approx$ 0.1\% on the
energy and $<$150 ps on the time (for high enough
signals such that the noise becomes negligible).
 Three time samples centred around the maximum are
used to perform this digital filter computation. This
is a compromise between noise reduction and sensitivity
to accidental showers. The sample immediately after
a gain change is not well measured and in a small fraction
of the cases only two samples can be used to reconstruct
energy and time.

\section{Shower reconstruction}

\subsection{Shower size}

 In this quasi-homogeneous calorimeter, with a Moliere
radius of $\approx$ 6.1 cm, the dominant intrinsic fluctuations in
the energy measurement come from fluctuations in the
energy loss outside the finite calorimeter area used to
measure the shower. 
To illustrate the importance of this effect, the
expected resolution from the GEANT 3.21 \cite{ref_geant}
Monte-Carlo program can be computed
for various dimensions of the area used to collect
the energy. For an infinite size, the ``sampling'' term
would be $\approx$ 1.2\%/$\sqrt{E}$. Using a R=11 cm radius,
this increases to $\approx$ 2.8\%/$\sqrt{E}$, and
$\approx$ 3.5\%/$\sqrt{E}$ for a 7 cm radius.
 The size used is a compromise between the electronic
noise (which increases with R) and the sampling term (which
decreases with R). The value used is R=11 cm, which contains
$\approx$ 100 calorimeter cells. 
This value is kept constant
with energy in order not to bias the linearity.
Only the most energetic
central cells are used for the time measurement, and the
time of all the other cells is fixed to this time in the
digital filter computation, again to avoid bias on the
linearity.
 In events with several showers, some care has to be taken
to subtract the energy leakage from one shower to another,
up to shower separation as large as $\approx$ 60 cm, to eliminate
small systematic biases in the energy measurement.
 The measurement of the impact point of the electron or photon
is based on the centre of gravity of the shower measured
in 3$\times$3 cells, to benefit from the narrow central core
of the shower.


\subsection{Position resolution}

\begin {figure}[ht]
\begin{center}
\includegraphics*[scale=0.40]{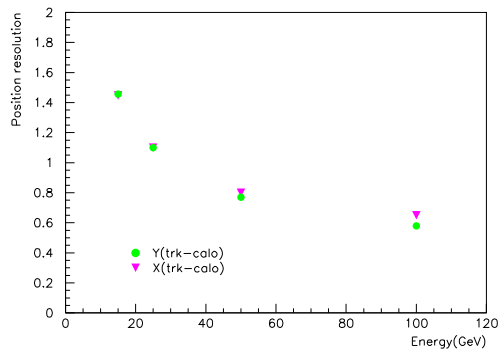}
\caption{\it Position resolution (in mm) as a function of
energy, measured with an electron beam.}
\label {fig_sigmaxyvse}
\end{center}
\end {figure}

 The position resolution has been studied using
data taken with a monochromatic
electron beam in 1996, by comparing the measured electron
shower position to the extrapolated track impact point
given by the NA48 spectrometer. The resolution of the later
quantity is $\approx$ 100$\mu$m and is negligible compared
to the calorimeter resolution. The results achieved are
shown in Figure \ref{fig_sigmaxyvse}, as a function
of the electron energy, for
both the $x$ (horizontal) and $y$ (vertical)
coordinates. The position resolution
is better than 1~mm above 25 GeV.

\subsection{Time resolution}

 The photon time resolution can be studied using
events of the type $K \rightarrow 3\pi^0 \rightarrow
(n) \gamma e^+ e^-$, by comparing the photon time to
the time of the electron tracks (given by a scintillator
array in front of the calorimeter). The time resolution
per photon is better than 500 ps.  
 In $K \rightarrow \pi^0 \pi^0$ events, the final event
time is computed combining the time informations from
the four photons. The event time resolution is better
than 250 ps. The tails outside $\pm$2~ns are
smaller than $10^{-4}$ level, which is a crucial
point for the validity of the $K_S$/$K_L$ identification method
used by NA48 \cite{ref_sabine}.

\subsection{Uniformity and energy resolution}

\subsubsection{\it Ke3 sample}

 The main tool to study in situ the detailed
performances of the calorimeter is given by a sample
of $K_L \rightarrow \pi^{\pm} e^{\mp} \nu$ decays
(called Ke3 decays), which
is very abundant. The electron track can
be measured in the spectrometer upstream of the
calorimeter. This spectrometer \cite{ref_spectro} consists of four
large drift chambers with a magnet between the
second and third chambers (which gives a $p_t$ kick
of 265 MeV/c). The position resolution of the drift
chambers is better than 100$\mu$m. This allows to
reconstruct the electron momentum ($p$) with a resolution
given by :
\begin{equation}
\sigma(p)/p = 0.5\% \oplus (0.009 \% \times p) \nonumber
\end{equation}
where $p$ is expressed in GeV.
 This measurement of $p$ can be compared
to the electron energy $E$ measured with the calorimeter.
In an ideal world, the ratio $\frac{E}{p}$ should be 1.
Taking p as ``perfect'', this allows to study the
variations in the energy response, the uniformity of
the response over the full calorimeter, the energy
resolution and the linearity.
A total of $\approx$ 150$\times$10$^6$ Ke3 decays
were recorded in 1998 and 1999 for this purpose.

\subsubsection{\it Response variation within one cell}

\begin {figure}[ht]
\begin{center}
\parbox{0.45\linewidth}
{
\includegraphics*[scale=0.30]{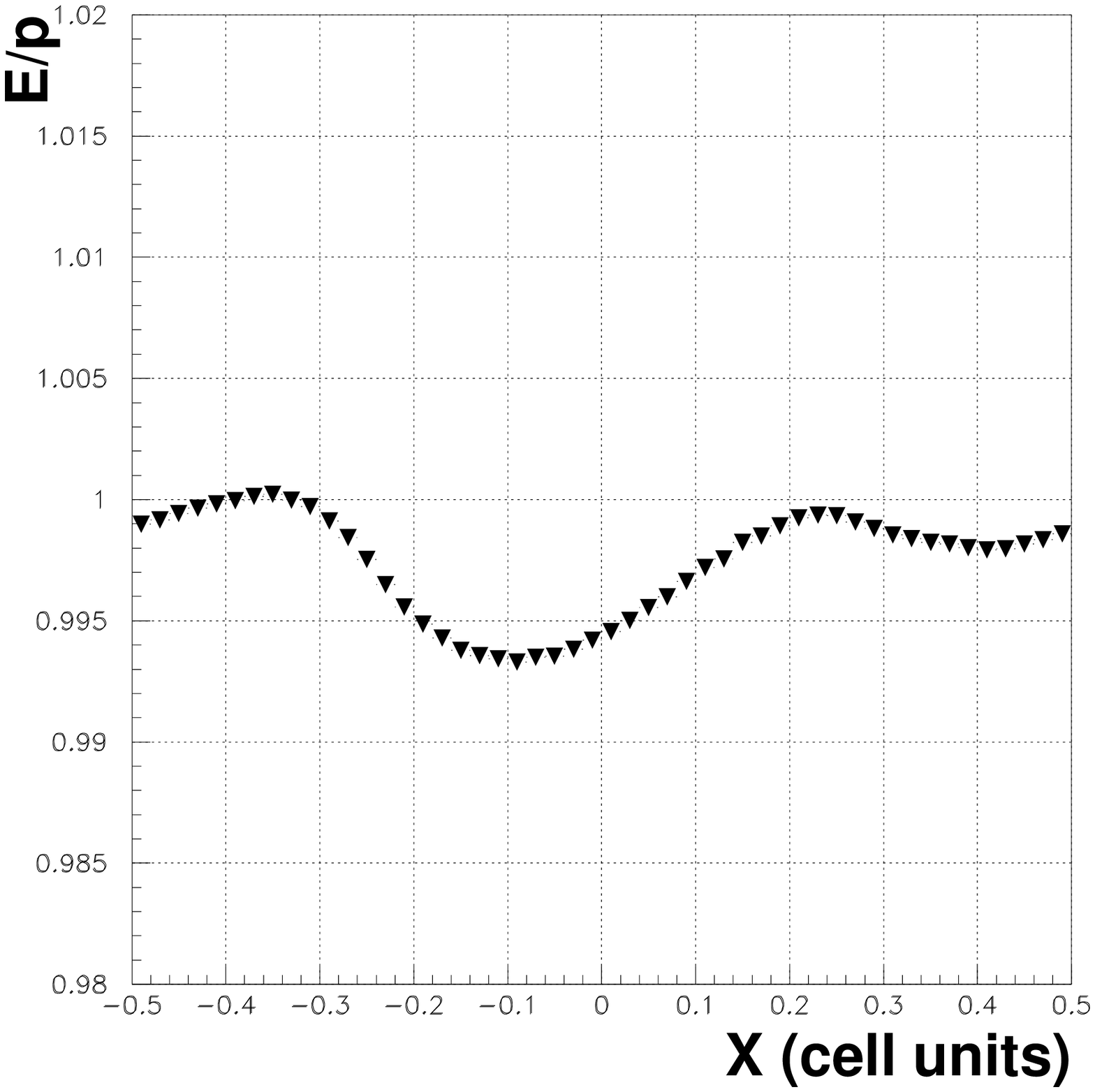}
}
\parbox{0.45\linewidth}
{
\includegraphics*[scale=0.30]{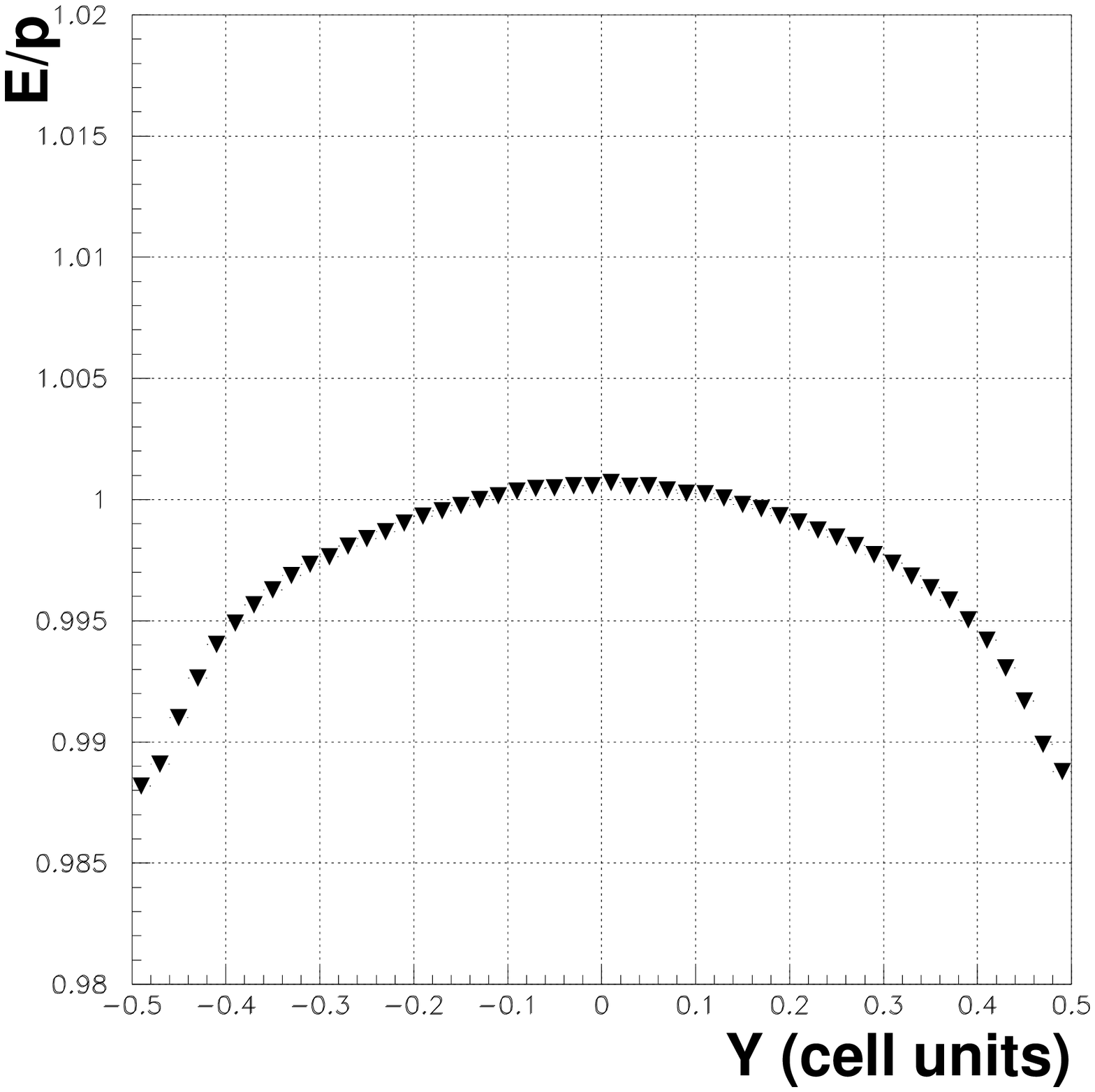}
}
\caption{\it Variation of E/p as a function of the
impact point within the cell in the horizontal (left)
and vertical (right) directions.}
\label {fig_eop_xcell}
\end{center}
\end {figure}

 Figures \ref{fig_eop_xcell} 
show the variation of $\frac{E}{p}$ as a function
of the impact point of the electron within one cell.
 The $\approx$ 0.5\% variation with $x$ (horizontal
direction) comes from the fact that when the electron
shower develops near the anode ($x$ $\approx$ 0),
the response is slightly reduced because part of
the deposited charge drifts during a time smaller than
the $\approx$ 70~ns integration time of the shaping.
Thanks to the accordion angle,
this effect is small and has a smooth variation with
the impact point, it can therefore be easily corrected.
 The drop of the response ($\approx$ 1\%) near the vertical separations
between cells comes from the fact that in the small
vertical gap between two anodes, the electric field
is reduced. This variation can also be corrected using
the measured shower position.
 Both these variations can be reproduced by a Monte-Carlo
which includes a detailed electric field map, as well
as a simulation of the readout electronics.
 After corrections, the residual variations with the
impact point within the cell are $\approx$~0.1\%.

\subsubsection{\it Intercalibration}

 The uniformity of the response over the full calorimeter
is studied by computing the average value of $\frac{E}{p}$ 
per cell (of the maximum energy deposition of the shower).
Using only the electronic calibration, and the values
of $\kappa$ measured as described above, the RMS dispersion
of $<\frac{E}{p}>$ is 0.41\%. This is well consistent with
the expected accuracy in the measurement of $\kappa$ 
of $\approx$ 1\%, given the fact that $\le$~40\% of
the shower energy is deposited in the impact cell.
 To improve the uniformity, Ke3 events are used
to equalise $<\frac{E}{p}>$ for all impact cells. A simple
iterative procedure is used to obtain cell per cell
coefficients \cite{ref_jose}. For this procedure,
only electrons in the energy range 25 to 40 GeV are used,
to avoid possible interplay between non-linearity and
non-uniformity.
This gives intercalibration factors which are
one static factor per cell, equivalent to
a redefinition of the $\kappa$ constants. Applying to the
99 Ke3 data the intercalibration factors derived from 
the 98 Ke3 sample, the RMS dispersion of $<\frac{E}{p}>$
as a function of the impact cell becomes $\approx$ 0.15\%,
showing the stability of the intercalibration procedure
(part of this dispersion is directly coming from the
event statistic used in 98 to derive the intercalibration
factors).
The validity of this procedure is also checked using
photons produced in $\pi^0$ and $\eta$ decays, produced
with $\pi^-$ beam striking thin targets during special
runs. The long range uniformity is found to be
better than $\approx$~0.1\%.

\subsubsection{\it Energy resolution}

\begin {figure}[ht]
\begin{center}
\includegraphics*[scale=0.37]{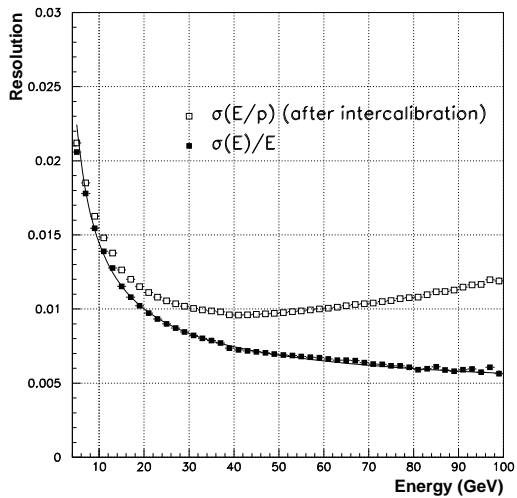}
\caption{\it Resolution on E/p and on E, after unfolding
the momentum resolution, as a function of the electron
energy, using the Ke3 sample collected in 99.}
\label {fig_resolution_ke3_99}
\end{center}
\end {figure}

 From the observed $\frac{E}{p}$ resolution, the
energy resolution can be deduced unfolding
the momentum resolution. Figure \ref{fig_resolution_ke3_99} shows
the $\frac{E}{p}$ and $E$ resolutions in the Ke3 sample
collected in 99, after the intercalibration procedure
described above. Because of the different
zero suppression scheme used for Ke3 events and
$K \rightarrow \pi^0\pi^0$ events, the energy measurement
in Ke3 events is based on 7$\times$7 calorimeter cells.

The corresponding energy resolution extrapolated
to the full shower size used for photons is :
\begin{equation}
\sigma(E)/E = (3.2\pm0.2)\%/\sqrt{E} \oplus (0.09\pm0.01)/E \oplus (0.42\pm0.05)\% \nonumber
\end{equation}
where E is in GeV. The uncertainties quoted include
uncertainties in the momentum resolution and in the
extrapolation to the full shower size.
Above 20~GeV, the energy resolution is better than 1\%. 
At the average energy of 25~GeV, the dominant term
in the energy resolution is the sampling term.
The coherent noise contribution to the total noise
of $\approx$ 90 MeV is almost negligible. 
 
 Several effects are expected to contribute to the
observed constant term~: a GEANT based
simulation gives a constant term of $\approx$ 0.2\%
(residual impact point variations, fluctuations in
longitudinal shower development, ...); the
intercalibration accuracy is $\approx$ 0.15\%,
the gap size variations ($\pm$45 $\mu$m) should
give a constant term contribution of $\approx$ 0.1 to 0.2\%;
the pulse reconstruction has a $\approx$ 0.1\% accuracy in
calibration events, and because the physics signal
shape is slightly different from calibration, one
expects an additional $\approx$ 0.15\% constant term
contribution. These effects can account qualitatively for
the constant term observed in the data, and this shows that
there is not one single dominant contribution to the
residual constant term.
If one did not use the Ke3 intercalibration, the
constant term would increase to $\approx$ 0.6\%. 

 The resolution quoted above is the result of a gaussian
fit. Some non-gaussian tails are present in the
energy response : $\approx$ 1\% of the showers have
a measured energy lower by more than 3$\sigma$ from the
average value, with $\approx$ 0.1\% having a response
20\% or more lower. This tail can be interpreted as
coming from $\pi^{\pm}$ production in electromagnetic
showers through hadronic photoproduction.

\subsection{Linearity}

\begin {figure}[ht]
\begin{center}
\includegraphics*[scale=0.37]{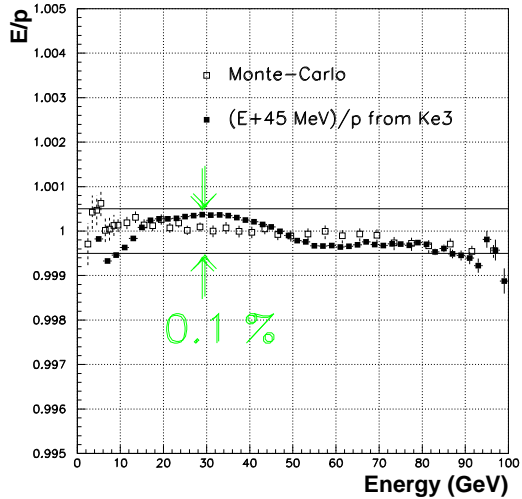}
\caption{\it Average value of E/p as a function of the energy,
showing the linearity of the response, for the Ke3 sample
collected in 99.}
\label {fig_linearity_ke3_99}
\end{center}
\end {figure}

 The average value of $\frac{E}{p}$ as a function
of the electron momentum is shown in
Figure \ref{fig_linearity_ke3_99}, where
45 MeV has been added to the electron energy to
account for energy loss before the Liquid Krypton.
 The energy response is linear in the energy range
5-100 GeV within $\approx$ 0.1\%. This figure
also shows the expectation from Monte-Carlo, which has
a small $\approx$ 0.05\% non linearity coming
from the gap opening~: as the energy increases,
the shower develops later in the calorimeter at a
place where the gap is larger and therefore the
initial current smaller.
 The small structure visible in the data at $E$
$\approx$ 30 to 50 GeV is probably coming from
a small non linearity in the ADC.

\section{$K \rightarrow \pi^0 \pi^0$ events and overall energy scale}

 The resolution of the $\pi^0$ mass in $K \rightarrow \pi^0 \pi^0$ events
observed in the data is $\approx$ 1 MeV, consistent with
the expectations based on the energy and position resolutions
described above. Thanks to this very good resolution, the
level of residual $K \rightarrow 3\pi^0$ background is smaller
than 0.1\% and induces a negligible systematic uncertainty in
the measurement of $\epsilon'/\epsilon$.

 As discussed in the introduction, the decay region definition
relies on the measured photon energies and positions:
the distance between the reconstructed decay vertex position
and the calorimeter is for instance directly
proportional to the calorimeter energy scale. Systematic
effects from non linearity are small as shown from the
Ke3 analysis. One is
then only left with fixing the overall energy scale factor of
the calorimeter. Given the small difference in electron/photon
response from the dead matter before the calorimeter, this
is done using the $K \rightarrow \pi^0 \pi^0$ events themselves.
In the $K_S$ beam, an anticounter located at the beginning of the
decay region (122~m upstream of the calorimeter)
is used to sharply define its beginning, by vetoing decays
occurring upstream.
The calorimeter energy scale is adjusted such that
the corresponding sharp rise in the reconstructed decay vertex
position matches the known geometrical position of this counter
with respect to the calorimeter. This is done with an
accuracy of few 10$^{-4}$. As a cross-check of non linearity,
one can check that this factor is constant with the kaon
energy (in the range 70- \linebreak 170 GeV) to better than $5\times10^{-4}$.

\section{Conclusions}

 The performances of the NA48 liquid Krypton
calorimeter have been studied extensively in situ.
The resolutions achieved ($\approx$ 500 ps time
resolution per photon, better than 1~mm position
resolution above 25~GeV, better than 1\% energy
resolution above 20~GeV with a constant term
smaller than 0.5\%), together with
the small non linearity in the
energy response and the good stability of the
calorimeter over several years, match the
requirements for a precise measurement of
direct CP violation in the neutral kaon system.

\end{document}